\documentclass[a4paper,11pt,hyper]{JHEP3}
\usepackage{amsfonts,latexsym,graphicx,epsfig,amssymb,amsmath,mathrsfs}

\newcommand{\ka}{K\"ahler~}
\newcommand{\agt}{\hspace{0.3em}\raisebox{0.4ex}{$>$}\hspace{-0.75em}\raisebox{-.7ex}{$\sim$}\hspace{0.3em}}
\newcommand{\alt}{\hspace{0.3em}\raisebox{0.4ex}{$<$}\hspace{-0.75em}\raisebox{-.7ex}{$\sim$}\hspace{0.3em}}

\title{Modular invariant inflation}
\author{Tatsuo Kobayashi$^a$, Daisuke Nitta$^b$, Yuko Urakawa$^b$\\
a.~Department of Physics, Hokkaido University, Sapporo, 060-0810 Japan\\
b.~Department of Physics and Astrophysics, Nagoya University, Chikusa,
Nagoya 464-8602, Japan}

\abstract{Modular invariance is a striking symmetry in string theory,
which may keep stringy corrections under control. In this paper, we
investigate a phenomenological consequence of the modular invariance,
assuming that this symmetry is preserved as well as in a four dimensional (4D) low
energy effective field theory. As a concrete setup, we consider a
modulus field $T$ whose contribution in the 4D effective field theory
remains invariant under the modular transformation and study inflation drived
by $T$. The modular invariance restricts a possible form of
the scalar potenntial. As a result, large field models of
inflation are hardly realized. Meanwhile, a small field
model of inflation can be still accomodated in this restricted setup. The scalar potential traced during the slow-roll
inflation mimics the hilltop potential $V_{ht}$, but it also has a
non-negligible deviation from $V_{ht}$. Detecting the primordial
gravitational waves predicted in this model is rather challenging. Yet,
we argue that it may be still possible to falsify this model by
combining the information in the reheating process which can be determined
self-completely in this setup.}

\keywords{Moduli inflation, Modular invariance, Primordial perturbations}
\preprint{EPHOU-16-003}

\maketitle

\begin{document}


\section{Introduction}

Moduli fields $T$, which ubiquitously appear in superstring theory on six-dimensional
(6D) compact space, play an important role in string phenomenology and cosmology.
Their vacuum expectation values (VEVs) determine the size and shape of the 6D compact space.
Moduli fields also appear in four-dimensional (4D) low-energy effective
field theory, where their VEVs determine couplings such as gauge
couplings, Yukawa couplings, higher order couplings
\cite{Hamidi:1986vh,Cvetic:2003ch,Cremades:2004wa} and so on. (See for a review, e.g. Ref.~\cite{Ibanez:2012zz}.)

When the radius of a cycle of the 6D compact space, $R$, is much bigger than the string
scale, i.e.,  $R^2 \gg \alpha'$, stringy corrections stay small and we
can calculate their contributions in the 4D theory based on 
a classical geometrical description. 
On the other hand, for $R^2 \simeq \alpha'$, where
the stringy corrections become large, multi instanton effects would be important
and it is, in general, difficult to compute them. 
However, superstring theory leads a very unique property, the
T-duality. A theory with $R$ would be equivalent to another with
$1/R$. For instance, the 4D low energy effective field theory of
heterotic string theory on torus and orbifold compactifications at the
tree level is invariant under the T-duality. The
parameter change $R \to 1/R$ is a subgroup of the modular transformation
of complex moduli parameter, which will be shown explicitly.
Thus, in order to investigate the stringy corrections, including also
the case with a smaller $R$, one may want to impose the modular invariance.

Stringy one-loop corrections have definite properties under modular
transformation~\cite{Kaplunovsky:1987rp,Dixon:1990pc,Derendinger:1991hq}. 
Contributions of the moduli fields to the 4D low energy effective action
which originate, e.g., from multi (world-sheet) instanton effects and  
infinite tower of Kaluza-Klein modes are described by a $SL(2,{\bf Z})$ modular
function, which is a meromorphic function. 
Indeed, the $T$ dependencies thus appear in Yukawa couplings
and higher order couplings~\cite{Cremades:2004wa}, threshold corrections
on gauge couplings \cite{Kaplunovsky:1987rp,Dixon:1990pc,Derendinger:1991hq},  
and other perturbative and non-perturbative corrections  are
given by modular functions. (See for a review,
e.g. Ref.~\cite{Ibanez:2012zz}.) Ferrara et al. derived, in Ref.~\cite{Ferrara:1990ei}, 
an ansatz of the 4D effective Lagrangian which is modular invariant, including non-perturbative effects, 
in order to study moduli stabilization and supersymmetry (SUSY)
breaking. (See, for a generalization, Ref.~\cite{Cvetic:1991qm}.)

Whether the modular invariance is kept inside of a 4D effective field theory or not depends on a
way of the compactification. However, since the modular
invariance is a striking symmetry in string theory, taking it for
granted, in this paper, we study a phenomenological consequence, in case
the modular invariance is preserved in the 4D low energy effective field
theory, in particular, by considering the period of the cosmological
inflation.

As a concrete setup, we assume the presence of a modulus field which
contributes to the 4D effective field theory in such a way that the
modular invariance is preserved. Applying
the studies  in Refs.~\cite{Ferrara:1990ei,Cvetic:1991qm}  to an inflationary setup, we ask the following questions:
\begin{itemize}
 \item Can a modulus field $T$ which preserves the modular invariance drive
       a successful inflation, which is compatible with cosmological
       observations?  
 \item If yes, is there a distinguishable aspect in the presence of the modular invariance?
\end{itemize}
To be more explicit, we consider a 4D low energy effective action which
remains invariant when we change the modulus field $T$ under the
$SL(2,{\bf Z})$ modular transformation and investigate whether $T$ can lead to a successful inflation or not. 
To keep the generality, we do not presume any concrete models of string
theory and compactification mechanisms, which may lead to the 4D effective
field theory with the modular invariance.

To address an impact of the stringy corrections, in
Ref.~\cite{Abe:2014xja}, Abe et al. studied an inflation model where the
scalar potential is given by the Dedekind eta function, which is in the
modular form. (See also Refs.~\cite{Higaki:2015kta,Kappl:2015esy}.) The
scalar potential in this model is given by the one for the natural inflation
with a small modulation, which adds a tiny oscillatory feature to the
potential of the natural inflation. This small modulation may 
lead to a phenomenologically interesting feature; a primordial spectrum with a
detectable running or the one which may explain the deficit of the CMB
temperature power spectrum at $l = 20-40$~\cite{Planck15}. Since the modulation is typically
suppressed by $e^{- 2 \pi T_R}$, where $T_R$ is the real part of the
modulus field and describes the size of the 6D compact space, in order
to have a non-negligible contribution from the modulation, one may want to consider a smaller value of
$T_R$. However, the analysis in Refs.~\cite{Abe:2014xja, Higaki:2015kta,Kappl:2015esy} 
cannot be straightforwardly extended to study the case $T_R \lesssim 1$,
because in this case, the neglected stringy corrections can alter the theory in 4D. This may
also motivate us to consider an inflation model where the stringy
corrections are kept under control.

In the absence of the modular invariance in 4D, it is known that the
imaginary part of a modulus field, so called axion, can serve a good
candidate of the inflaton. At a perturbative level, axions have shift
symmetries, while they may be broken into discrete symmetries due to
non-perturbative effects such as world-sheet instanton effects. As a
consequence, a potential of axions can be generated. Usually, the
inflation driven by axions is periodic like the natural inflation~\cite{Freese:1990rb}
(for a review of the axion inflation, see e.g.,
\cite{Pajer:2013fsa}). In such a case, in order to maintain a slow-roll evolution
sufficiently long, the axion decay constant $f$ needs
to be larger than the Planck scale $M_{pl}$. The recent Planck 
measurement \cite{Planck15} puts the lower bound on the decay constant
in the natural inflation as $\log_{10} (f/M_{pl})> 0.84$ at 95\% CL. 
However, it was suggested that controlled axion inflation models
constructed in string theory may generically predict $f < M_{pl}$~ \cite{Svrcek:2006yi}. This difficulty has
been challenged and it was shown that (even if a
bare value of the decay constant is smaller than $M_{pl}$) an
effective value of the decay constant can be increased to be
compatible with the observations by considering an alignment of several
axions \cite{Kim:2004rp} and alternatively by considering one-loop effects
\cite{Abe:2014xja,Abe:2014pwa}.

In this paper, we show that an inflationary solution which successfully
ends can be accommodated in the low energy effective field theory with
the modular invariance. We name this model {\it modular invariant
inflation}. Since the modular invariance prohibits us to introduce a
constant term in the superpotential, the corresponding scalar potential
differs from the one studied in Ref.~\cite{Abe:2014xja}. A notable
aspect of the modular invariant inflation is that it is a small field
model and a slow-roll trajectory can be realized without an additional
machinery which increases the decay constant. In this paper, for
simplicity, we consider one modulus field $T$, which yields two independent
fields as $T$ is a complex field.

This paper is organized as follows. In Sec.~\ref{Sec:MI}, after giving a
brief explanation of the modular invariance, we describe our setup of
the modular invariant inflation. Then, we study in detail our modulus
potential, paying attention to several features which are implemented by
the modular invariance. In Sec.~\ref{Sec:Search}, after we give the
basic equations in this model, we explore an
inflationary solution and show that the slow-roll evolution can be
indeed taken place around the edge of the fundamental region in the $T$
space. In Sec.~\ref{Sec:Primordial}, we compute the primordial spectrums
and discuss the consistency with the Planck measurements. Section
\ref{Sec:Conclusion} is devoted to conclusion and discussion.

\section{Modular invariance in supergravity} \label{Sec:MI}
 In this paper, we consider a modulus field which remains invariant under the modular
transformation in the 4D low-energy effective action of superstring
theory. (See for a review, e.g. Ref.~\cite{Ibanez:2012zz}.) 
 For a notational brevity, we express chiral superfields and their
lowest components by the same letters. We omit the Planck mass, setting
$M^2_{\rm pl}= 8 \pi G =1$, unless necessary.

\subsection{\ka penitential and superpotential with modular invariance}
Using the \ka potential $K$ and the superpotential $W$, the scalar part of
the Lagrangian in supergravity is expressed as
\begin{align}
 & \frac{{\cal L}}{\sqrt{-g}} = - K_{I \bar{J}} \partial_\mu \Phi^I \partial^\mu \bar{\Phi}^{\bar{J}} - V(\Phi^I,\, \bar{\Phi}^{\bar{I}})\,,
\end{align}
where $K_I$ and $K_{I\bar{J}}$ are given by
\begin{align}
 & K_I \equiv \frac{\partial K}{\partial \Phi^I}\,, \qquad 
 K_{I \bar{J}} \equiv \frac{\partial^2 K}{\partial \Phi^I \partial \bar{\Phi}^{\bar{J}}}\,.  \label{Def:KI}  
\end{align}
The scalar potential in supergravity is given by 
\begin{align}
 & V= e^K \left[ K^{I \bar{J}} D_I W D_{\bar{J}} \bar{W} - 3 |W|^2 \right],
\end{align}
with
\begin{align}
 & D_I W \equiv K_I W + W_I \,.
\end{align}
Using the K\"ahler function $G$,
 \begin{align}
 & G = K + \ln W + \ln \bar{W} \,,
\end{align}
we can rewrite the scalar potential as
\begin{align}
 & V = e^G \left( G^{I \bar{J}} G_I G_{\bar{J}} - 3 \right) \,. 
\end{align}

We consider a model with a modulus $T$ and other chiral
superfields $X$ such as other moduli fields and matter fields.  We
do not impose the modular invariance on the superfields $X$~\footnote{In
Ref.~\cite{Copeland:1994vg}, the modular invariance was imposed on a waterfall field,
which drives the transition from the false vacuum to the true vacuum.}.
Their K\"ahler potential is written by 
\begin{align}
 & K = - n \ln (T + \bar{T}) + \kappa (X,\, \bar{X})\,. \label{Def:kahler}
\end{align}
Here, the parameter $n$ depends on properties of the modulus such as the
geometrical aspects. Typically, $n$ is $n=1,2,3$ or a fractional number.
For example, the overall K\"ahler modulus has $n=3$.
When the 6D compact space is $T^2 \times T^4$, 
the K\"ahler modulus corresponding to $T^2$ has $n=1$, while 
the volume modulus for $T^4$ has $n=2$.
In what follows, we focus on the case with $n=1,\, 2,\,3$.

Under the $SL(2,{\bf Z})$ modular transformation,
\begin{align}
 & T \to \frac{aT - i b}{i cT + d}  \label{MT}
\end{align}
with $ad-bc=1$ and $a,b,c,d \in {\bf Z}$, the kinetic term of $T$
derived from the \ka potential given in Eq.~(\ref{Def:kahler}):
\begin{equation}
K_{T \bar{T}} \partial_\mu T \partial^\mu \bar{T} = \frac{n}{(T +
\bar{T})^2} \partial_\mu T \partial^\mu \bar{T}
\end{equation}
remains invariant~\footnote{This kinetic term is 
invariant under $SL(2,{\bf R})$, but it is broken to the discrete symmetry $SL(2,{\bf Z})$ 
on the compact space, e.g. by world-sheet instanton effects.}.
The modular transformation is generated by two transformations:
\begin{align}
 & T \to \frac{1}{T}\,, \qquad T \to T + i \,,
\end{align}
and the invariance under the transformation is the stringy symmetry on
the compact space. In the following, we express $T$ as
\begin{align}
 & T = T_R + i T_I 
\end{align}
with $T_R >0$. Performing the modular transformation, one can map an arbitrary
point in the modulus space $T$ to a corresponding point within 
the fundamental domain, $|T| \geq 1$ and $-\frac12 \leq T_I \leq
\frac12$. 
\begin{figure}[t]
\begin{center}
\begin{tabular}{cc}
\includegraphics[width=7cm]{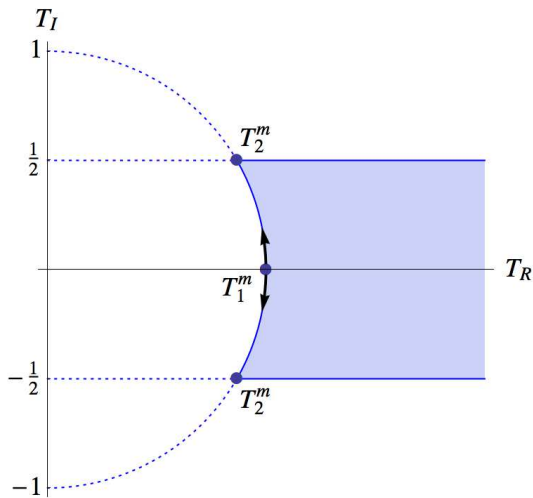}
\end{tabular}
\caption{The shaded region is the fundamental domain, to which all
 points in the $T$ space can be mapped by performing the modular
 transformation. We will find that the scalar potential $V$ has the
 extreme values at $T^m_1=1$ and $T^m_2 = e^{ \pm i \frac{\pi}{6}}$. The
 arrows denote the trajectories of the inflationary solution which
 will be found for $B> 0$.}
\label{Fg:FR}
\end{center}
\end{figure}
(See Fig.~\ref{Fg:FR}.)

To preserve the modular invariance of the whole Lagrangian, the scalar
potential should be also modular invariant. Since the \ka potential
changes under the modular transformation as 
\begin{align}
 & K \to K + \ln |i cT + d|^{2n}  \,,
\end{align}
the K\"ahler function $G$ and the scalar potential remain invariant if
the superpotential transforms as  
\begin{equation}
W \rightarrow (i cT + d)^{-n} W ,
\end{equation}
that is, the modular weight of $W$ should be $-n$. Using the the
Dedekind eta function:
\begin{align}
 & \eta(T) = e^{- \frac{\pi}{12} T} \prod_{n=1}^\infty (1- e^{- 2 \pi
 nT})  \qquad (T_R >0) \,,
\end{align}
which is in the modular form with the modular weight $1/2$, i.e., transforms as
\begin{align}
 & \eta(T) \to (icT + d)^{1/2} \eta(T)\,,
\end{align}
we can construct a superpotential $W(T)$ whose modular weight is $-n$ as
\begin{align}
 & W = \frac{\omega(X)}{\eta(T)^{2n}}\,. \label{Def:super}
\end{align}
Then, since the change of the \ka potential in $G$ is canceled by the
change of the superpotential, the scalar potential and the Lagrangian
density for the effective field theory in 4D are now guaranteed to be
modular invariant~\cite{Ferrara:1990ei,Cvetic:1991qm}.

Notice that the modular invariance prohibits to introduce
a constant term in the superpotential. In the absence of the constant
term, the cosine term, which introduces the deviation from the
cosmological constant in the natural inflation, does not appear in the
scalar potential. As a result, in the modular invariant inflation, we
obtain a qualitatively different scalar potential from the usual axion
inflation models with the natural inflation type potential (see, e.g.,
Ref.~\cite{Abe:2014xja}).

\subsection{Lagrangian and equations of motion}
The Lagrangian density is given by
\begin{align}
 & {\cal L} = - \sqrt{-g} \left[ \frac{n}{4 T_R^2}\, (\partial_\mu T_R
 \partial^\mu T_R + \partial_\mu T_I \partial^\mu T_I) + V(T_R,\, T_I)
 \right]. \label{Exp:Lag}
\end{align}
Taking the derivative of the Lagrangian density (\ref{Exp:Lag}) with respect to
$T_R$ and $T_I$, we obtain the equations of motion as
\begin{align}
 & \frac{1}{\sqrt{-g}} \partial_\mu \left( \frac{n}{2 T_R^2} \sqrt{-g}
 g^{\mu \nu} \partial_\nu T_R \right) + \frac{n}{2 T_R^3} \left(
 \partial_\mu T_R \partial^\mu T_R + \partial_\mu T_I \partial^\mu T_I
 \right) - V_R = 0\,, \\
 & \frac{1}{\sqrt{-g}} \partial_\mu \left( \frac{n}{2 T_R^2} \sqrt{-g}
 g^{\mu \nu} \partial_\nu T_I \right)- V_I = 0\,,
\end{align}
where we defined
\begin{align}
 & V_R \equiv \frac{\partial V}{\partial T_R}\,, \qquad V_I \equiv
 \frac{\partial V}{\partial T_I}\,.
\end{align}
As is usual in inflation models from supergravity, the fields $T_R$ and
$T_I$ have the non-canonical kinetic term.

\subsection{Scalar potential}   \label{SSec:V}
In our setup, we assume that the modulus fields $T_R$ and $T_I$ are the
inflaton fields and the other fields $X$ are stabilized to constant values during inflation. Using 
\begin{align}
 & D_T W = - n \left( \frac{1}{T + \bar{T}} + 2
 \frac{\eta_T(T)}{\eta(T)} \right) W \,, \\
 & D_X W = \left( \kappa_X + \frac{\omega_X}{\omega} \right) W\,, 
\end{align}
where $\eta_T$ denotes the derivative of $\eta(T)$ with respect to $T$,  
we can compute the scalar potential $V$ as
\begin{align}
 & V = \frac{A}{(T + \bar{T})^n} \frac{1}{|\eta(T)|^{4n}} \left[ n
 \left| 1 + 2 (T + \bar{T}) \frac{\eta_T(T)}{\eta(T)} \right|^2 + B
 \right] + \delta V \,  \label{Exp:potential}
\end{align}
with
\begin{align}
 & A \equiv e^{\kappa(X,\, \bar{X})} |\omega(X)|^2 \geq  0\,, \\
 & B \equiv \kappa^{X \bar{X}} \left| \kappa_X + \frac{\omega_X}{\omega}
 \right|^2 -3 \geq -3 \,.
\end{align}
Since the fields $X$ are stabilized, $A$ and $B$ stay constant and serve
free parameters. Here, we added a correction $\delta V$ to the tree-level supergravity scalar potential, 
which may appear because of explicit supersymmetry breaking 
and/or loop effects.\footnote{Loop effects can be sizable, when there
exist a sufficiently large number of modes,  e.g. in the hidden sector\cite{Choi:1994xg}.}
The correction $\delta V$ can be positive and negative and the value of
$\delta V$ depends on the detail of each model. In our setup, we assume that the
correction $\delta V$ is independent of $T$ \footnote{
The modular invariance is unbroken unless $T$ develops its VEV.} 
and $\delta V$ can take both
positive and negative values.

\begin{figure}[t]
\begin{center}
\begin{tabular}{cc}
\includegraphics[width=10cm]{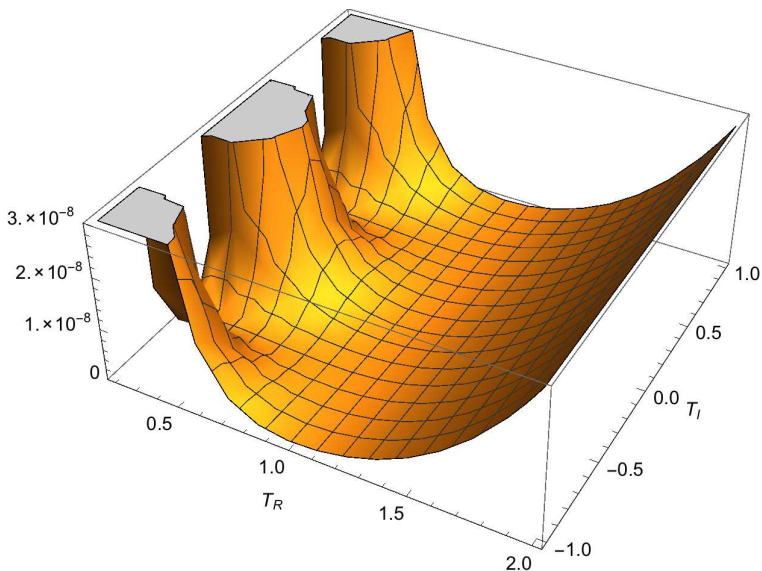}
\end{tabular}
\caption{Scalar potential for  $n=1$, $B=0.6$, and $A= 10^{-8}$.}
\label{Fg:V3D}
\end{center}
\end{figure}
The modular invariance implies that the values of $V$ at two points,
which are related by the modular transformation, should
coincide. Therefore, when $V$ at a point outside the fundamental domain
takes a particular value, we can find another point inside the
fundamental domain which has the same value of $V$. As one may expect
from this fact, the extreme values of the scalar potential can be found
on the edge of the fundamental domain. Among the points on the edge of
the fundamental region,
\begin{eqnarray}
  & T^m_1 \equiv 1 \,,
\end{eqnarray}
and
\begin{align}
 & T^m_2 \equiv \frac{1}{2} (\sqrt{3} \pm i) = e^{\pm i\frac{\pi}{6}}\,
\end{align}
are mapped into themselves under $T \to \frac{1}{T}$ with the
identification of $T$ and $T \pm i$. We find that the scalar potential
$V$ takes extreme values at these points $T^m_1$ and $T^m_2$. In
particular, for $B>0$, $T^m_2$ becomes the global minimum and for 
$-3 <  B < 0$, $T^m_1$ becomes the global minimum~\footnote{We
found that for $B \simeq -3$, the global minimum may be located at
another point near $T^m_1$. In the rest of this paper, we do not
discuss such a value of $B$, but the gradient of $V$ for these values of
$B$ is steep and hence it seems difficult to find a successful
inflationary solution}. For $B=0$, both $T^m_1$ and $T^m_2$ become the
global minimum. For the particular ansatz of $W$ and $K$, which ensures
the modular invariance, the absolute value in the
square brackets in Eq.~(\ref{Exp:potential}), 
i.e.,
$$
\left| 1 + 2 (T + \bar{T}) \frac{\eta_T(T)}{\eta(T)} \right|\,\, \left(  \propto   \left|\frac{ D_T W}{W} \right| \right)\,,
$$
vanishes at both $T^m_1$ and $T^m_2$. Therefore, setting $V=0$ at the global minimum, we find the relation between $\delta V$, $A$, and $B$ as 
\begin{align}
 & \delta V  =  - \frac{A B}{(T_a^m + \bar{T}_a^m)^n
 |\eta(T_a^m)|^{4n}}\,, 
 \label{Exp:delta V}
\end{align}
and obtain the scalar potential as
\begin{align}
 & V = \frac{A}{(T + \bar{T})^n} \frac{1}{|\eta(T)|^{4n}} \cr
 & \qquad  \times \left[ n
 \left| 1 + 2 (T + \bar{T}) \frac{\eta_T(T)}{\eta(T)} \right|^2 + B
 \left\{ 1 - \left( \frac{T+ \bar{T}}{T_a^m + \bar{T}^m_a} \right)^n \left|
 \frac{\eta(T)}{\eta(T^m_a)} \right|^{4n } \right\}
 \right]\,.  \label{Exp:potential2}
\end{align}
For $B>0$, the global minimum is located at $T^m_2$, i.e., $a=2$ and for $B<0$, it
is at $T^m_1$, i.e., $a=1$.

Notice that the scalar potential $V$ is determined solely by the three
parameters $A$, $B$, and $n=1,\,2,\,3$. The positive parameter $A$
determines the magnitude of the scalar potential and $B$ and $n$
determine the gradient of $V$. In particular, whether an inflationary evolution can
be realized or not is only determined by $B$ and $n$, since
$A$ can be absorbed into the normalization of the Hubble parameter (by
using the $e$-folding number as a time coordinate). 

The scalar potential shows various features depending on the value of
$B$ and also on the position in the $T$ space. In Fig.~\ref{Fg:V3D}, we plot the scalar potential $V(T_R,\, T_I)$ for
$n=1$, $B=0.6$, and $A= 10^{-8}$. Because of the invariance under $T \to T+ i$, the potential in the
direction of the axion $T_I$ becomes periodic. The scalar potential $V$
blows up in the limit $T_R \gg 1$ and in the limit $T_R \ll 1$ for some
range of $T_I$. As a consequence of the modular invariance, a region
with a gentle slope can be found around $|T| = 1$, which is the edge of the fundamental region (see
Fig.~\ref{Fg:FR}). This feature may become more obvious in the contour plot. 
\begin{figure}[t]
\begin{center}
\begin{tabular}{cc}
\includegraphics[width=7cm]{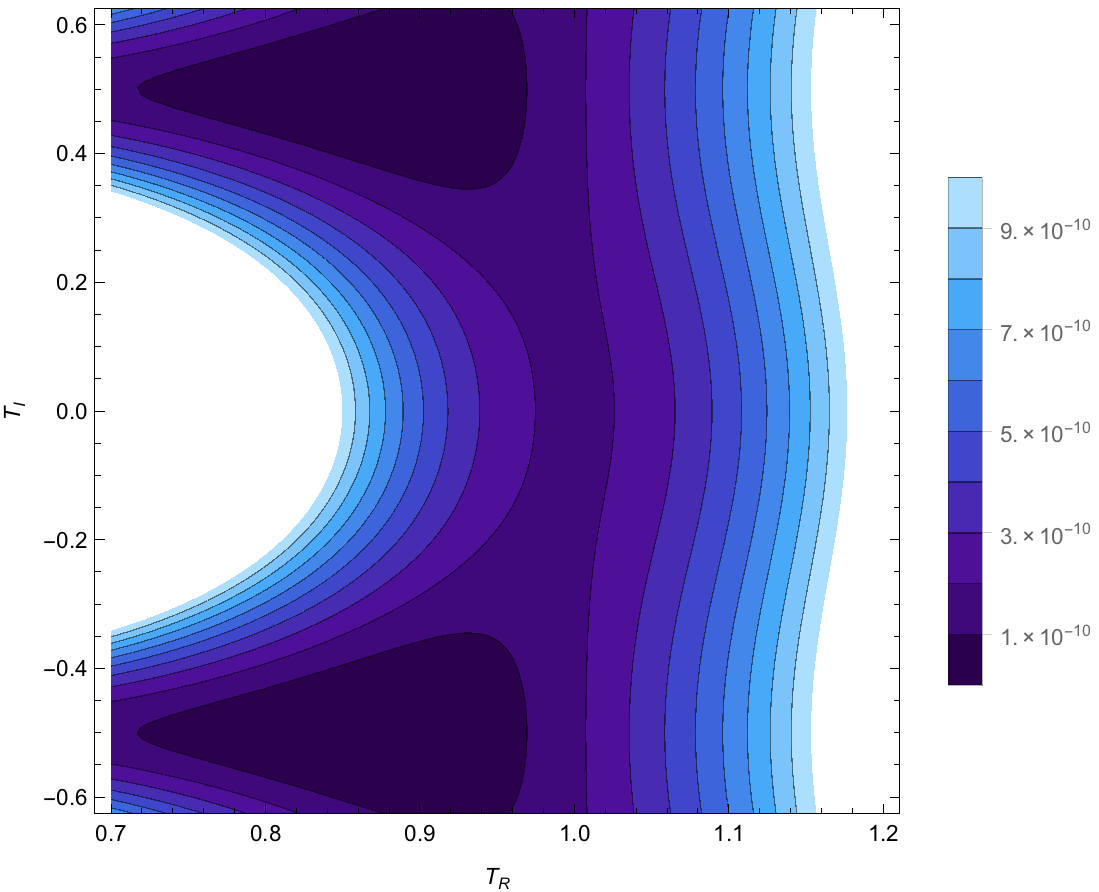}
\includegraphics[width=7cm]{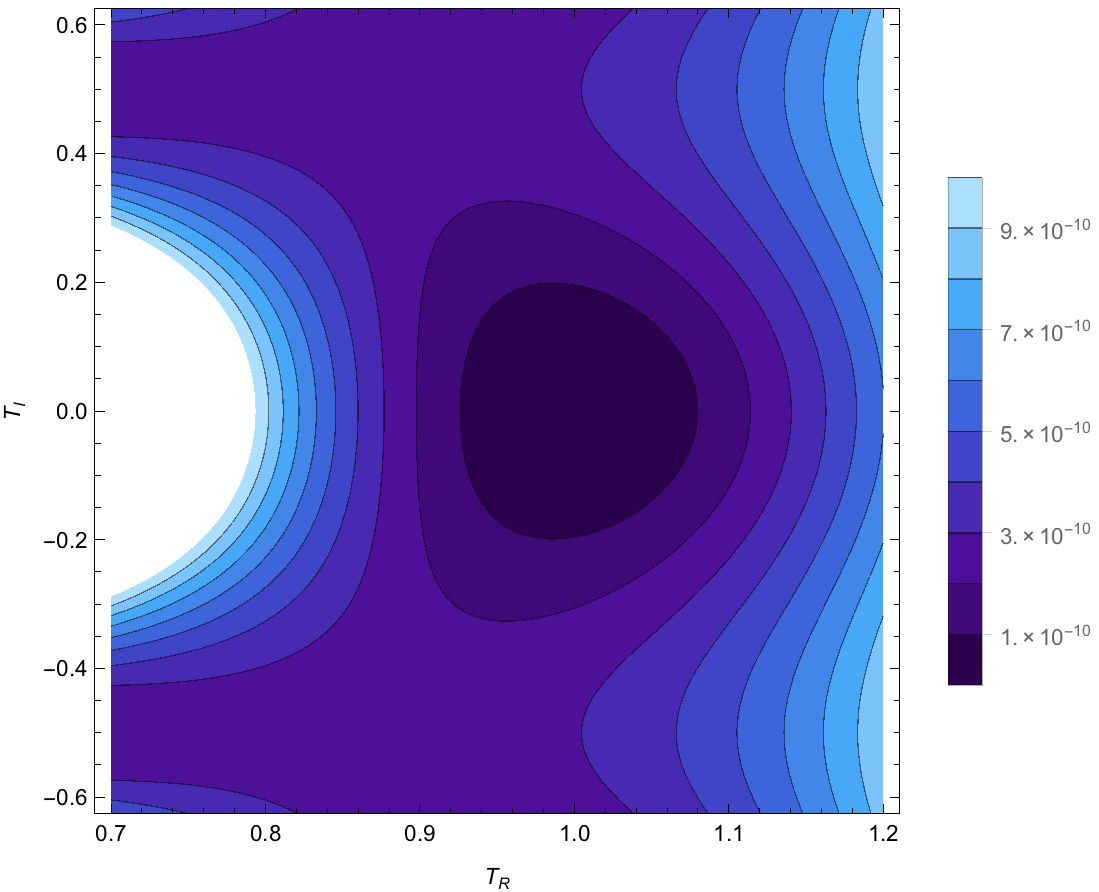}
\end{tabular}
\caption{The left panel shows the scalar potential $V$ for $n=1$ and
 $B=0.6$ and the right panel shows $V$ for $n=1$ and $B=-0.8$. A region
 with a darker/brighter color has a smaller/larger value of $V$. For
 $B>0$ (the left panel), the global minimum is located at $T^m_2 =
 e^{\pm i \pi/6}$ and
 for $B<0$ (the right panel), it is located at $T^m_1=1$. }
\label{Fg:VCP}
\end{center}
\end{figure}
In Fig.~\ref{Fg:VCP}, we show the contour plots of the scalar potential
$V$ for $A= 10^{-1}$ and $B=0.6$ (Left) and $B=-0.8$ (Right). For these values of $B$, indeed, the gradient of $V$
around $|T|=1$ becomes comparatively shallow.

\begin{figure}[t]
\begin{center}
\begin{tabular}{cc}
\includegraphics[width=7cm]{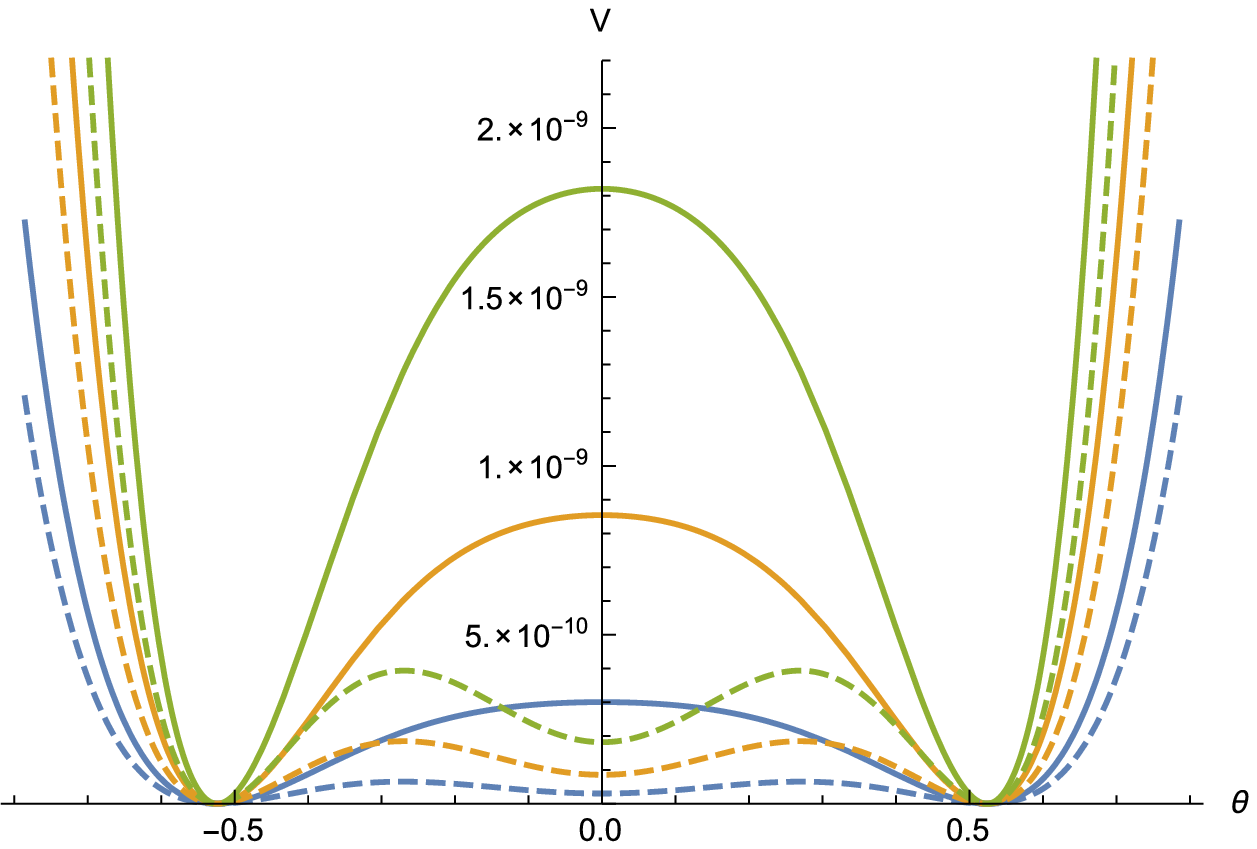}
\includegraphics[width=7cm]{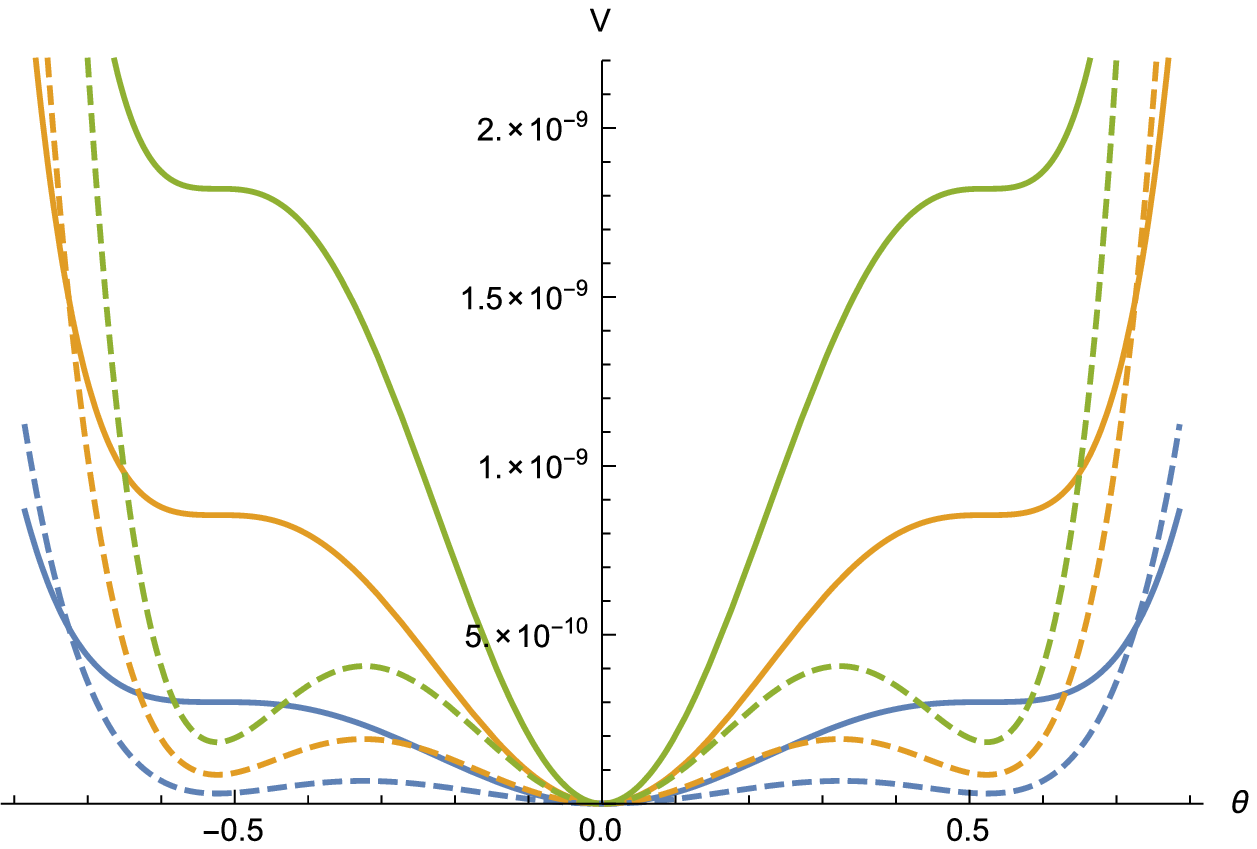}
\end{tabular}
\caption{The left panel shows the scalar potential $V$ for positive
 values of $B$ and the left panel shows the scalar potential $V$ for
 negative values of $B$. The blue lines are for $n=1$, the orange lines
 are for $n=2$, and the green lines are for $n=3$. The solid lines are
 for $B= \pm 1$ and the dashed lines are for $B= \pm 0.1$. We set
 $A=10^{-8}$. }
\label{Fg:Vtheta}
\end{center}
\end{figure}
To capture the potential feature in more detail, in Fig.~\ref{Fg:Vtheta}, we present the scalar potential on
$|T|=1$, i.e.,  $T=e^{i \theta}$. As described previously, the extreme values can be found at $\theta =0$ and
$\theta=\frac{\pi}{6}$.  For $B>0$, as we decrease $B$, the point
$\theta=0~(T^m_1)$ is transformed from the local maximum to the local minimum. In particular, the
potential around $\theta \simeq 0$ becomes almost flat for $B \simeq 0.6$.
Meanwhile, for $B<0$, $\theta= \frac{\pi}{6}~(T^m_2)$ is transformed from the inflection point
to the local minimum. The potential for $\frac{\pi}{7} \alt \theta \alt \frac{\pi}{6}$
becomes nearly flat for $B \simeq - 0.8$.  

\subsection{Gross feature of $V$: Hilltop potential}  \label{SSec:Gross}
\begin{figure}[t]
\begin{center}
\begin{tabular}{cc}
\includegraphics[width=7.5cm]{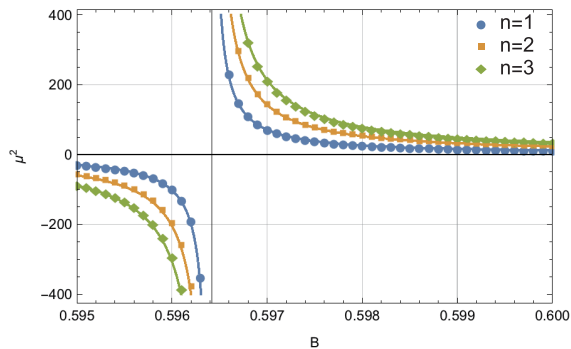}
\end{tabular}
\caption{This figure shows the $B$ dependence of $\mu^2$. The solid line shows the function $\mu^2(B)= 0.042\,n/(B- B_c)$.}
\label{Fg:Vht}
\end{center}
\end{figure}

To get a better intuition on the gross feature of the scalar potential
for $B \simeq 0.6$ on $|T| \simeq 1$, where $V$ becomes almost flat
namely around $T \simeq 1$, we
look for a simpler function which can approximate $V$ well. Expressing
$T$ as $T= e^{i \theta}$, we consider the scalar potential for 
$|\theta| \ll 1$. Then, the modular invariance, namely the invariance under $T \to 1/T$, requires
the invariance under the change $\theta \to - \theta$. Since $T_I \simeq \theta$ 
for $|\theta| \ll 1$, keeping only the terms which are invariant 
under $T_I \to - T_I$, we approximate the scalar potential as
\begin{align}
 & V_{ht}= \Lambda^4 \left[ 1- \left( \frac{T_I}{\mu} \right)^2 \right]\,, \label{Vht}
\end{align}
which agrees with the leading terms of the hilltop inflation during
inflation. The value of $\mu^2$ which well approximates $V$ for each
value of $B$ is presented in Fig.~\ref{Fg:Vht}.
Here, we determined $\Lambda$, setting $V=V_{ht}$ at $\theta =0$, and
determined $\mu$, setting $V=V_{ht}$ at $\theta = 10^{-3}$. As is shown
in Fig.~\ref{Fg:Vht}, the $B$ dependence of $\mu^2$ can be approximately 
given by the following function 
\begin{align}
 & \mu^2 (B)= \frac{0.042\times n}{B-B_c}\,, \qquad B_c \equiv 0.59641
 \cdots \, .  \label{tempmu}
\end{align} 
For $B \alt B_c$, the curvature of the potential in the direction
$T_I$ is positive and $\theta=0$ becomes the local minimum. Meanwhile,
for $B \agt B_c$, the curvature is negative and $\theta=0$
becomes the local maximum. This suggests that for $B \agt B_c$ we may
find an inflationary solution which ends successfully. This will be
verified in the next section.

\section{Exploring inflationary solution}  \label{Sec:Search}
In this section, after we give the background equations of motion, we
explore an inflationary solution with a sufficiently large $e$-folding
number.  

\subsection{Background evolution and slow-roll parameters}
In the Friedmann-Robertson-Walker background, the field equations for $T_R$ and $T_I$ are given by
\begin{align}
 & \ddot{T}_R + 3 H \dot{T}_R + \frac{1}{T_R} (\dot{T}_I^2 -
 \dot{T}_R^2) + \frac{2}{n} T_R^2 V_R = 0\,, \label{Eq:TRbg} \\
 &  \ddot{T}_I + 3 H \dot{T}_I - 2 \frac{\dot{T}_R}{T_R} \dot{T}_I +
 \frac{2}{n} T_R^2 V_I = 0\,, \label{Eq:TIbg}
\end{align}
where $H$ is the Hubble parameter. As an action for gravity, we assume the Einstein-Hilbert action. Then,
the Hamiltonian constraint equation gives the Friedmann equation:
\begin{align}
 & H^2 = \frac{1}{3} \left[ \frac{n}{4 T_R^2} (\dot{T}_R^2 +
 \dot{T}_I^2) + V \right]\,.
\end{align}
Taking the time derivative of the Friedmann equation and using
Eqs.~(\ref{Eq:TRbg}) and (\ref{Eq:TIbg}), we obtain
\begin{align}
 & \dot{H} = - \frac{n}{4 T_R^2} (\dot{T}_R^2 + \dot{T}_I^2)\,. 
\end{align}

We introduce the slow-roll parameters for $T_R$ and $T_I$ as
\begin{align}
 & \varepsilon_{1,\alpha} \equiv \frac{n}{4 T_R^2} \left(
 \frac{\dot{T}_\alpha}{H} \right)^2 \,, \qquad  \varepsilon_{m+1, \alpha} \equiv \frac{\dot{\varepsilon}_{m,
 \alpha}}{H \varepsilon_{m, \alpha}} \qquad ({\rm for} \,\, m\geq 1)\,,
\end{align}
with $\alpha=R,\, I$. We also introduce the slow-roll parameter or the Hubble flow functions: 
\begin{align}
 & \varepsilon_1 \equiv - \frac{\dot{H}}{H^2}\,,  \qquad 
 \varepsilon_{m+1} \equiv \frac{\dot{\varepsilon}_m}{H \varepsilon_m}
 \qquad ({\rm for} \,\, m\geq 1) \,.
\end{align}
The slow-roll parameters $\varepsilon_1$ and $\varepsilon_{1,\, \alpha}$
are related as
$$
 \varepsilon_1 = \varepsilon_{1,R} + \varepsilon_{1, I}\,.
$$
Using the slow-roll parameters, we obtain
\begin{align}
 & \frac{\ddot{T}_\alpha}{H \dot{T}_\alpha} = \frac{1}{2} \varepsilon_{2,\alpha} +
 2 \sqrt{\frac{\varepsilon_{1, R}}{n}} - \varepsilon_1\,.
\end{align}
When the magnitude of the slow-roll parameters stay small, i.e., 
$|\varepsilon_{m, \alpha}| \ll 1$, the field equations are given by
\begin{align}
 & H^2 \simeq \frac{1}{3} V\,, \\
 & \frac{\dot{T}_R}{H} \simeq - \frac{2 T_R^2}{n} \frac{V_R}{V} -
 \frac{4 T_R^3}{3 n^2} \left( \frac{V_I}{V} \right)^2 \,, \label{Eq:dTRap} \\
 & \frac{\dot{T}_I}{H} \simeq  - \frac{2 T_R^2}{n} \frac{V_I}{V}\,. \label{Eq:dTIap}
\end{align}

\subsection{Case studies}
We investigate an inflationary solution for $B> 0$ in
Sec.~\ref{SSSec:Bp} and for $- 3 < B \leq 0$ in Sec.~\ref{SSSec:Bnp}.

\subsubsection{Case1: $B> 0$}  \label{SSSec:Bp}
In Sec.~\ref{SSec:Gross}, we examined the gross feature of the scalar
potential around $T \simeq 1$. As is suggested there, for
$B_c < B \alt 0.6$, we indeed find an
inflationary solution, which successfully ends. For $B \agt 0.6$, the
gradient of the scalar potential is too steep to realize an inflationary
evolution. For $B< B_c$, where the effective mass becomes positive, the
trajectory is either not realizing inflation (with a larger initial
velocity) or approaching to the eternal de Sitter, being settled at the
local maximum $T^m_1=1$ (with a smaller initial velocity).

\begin{figure}[t]
\begin{center}
\begin{tabular}{cc}
\includegraphics[width=7.5cm]{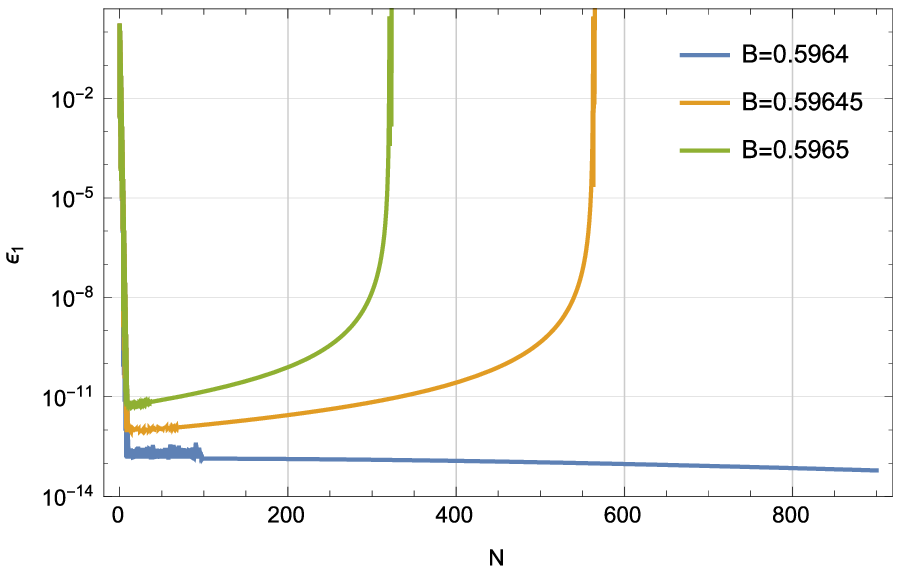}
\includegraphics[width=7.5cm]{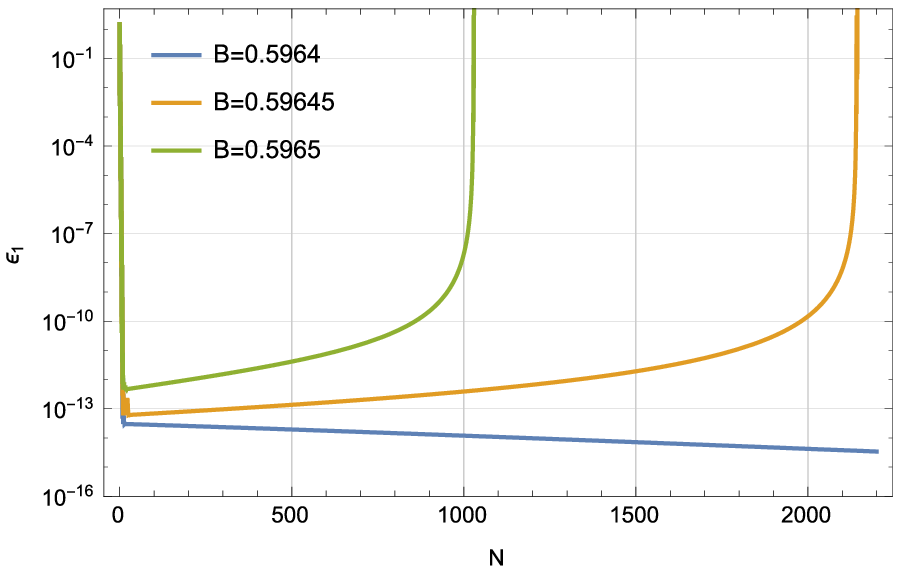}
\end{tabular}
\caption{We plot the time variation of the slow-roll parameter
 $\varepsilon_1$ for $n=1$ and $n=2$, setting $B$ to several different
 values. The blue line is for $B=0.5964 (< B_c)$, the orange line is for
 $B=0.59645 (> B_c)$, and the green line is for $B=0.5965 (> B_c)$.}
\label{Fg:epsilon}
\end{center}
\end{figure}
In Fig.~\ref{Fg:epsilon}, we plot the time evolution of $\varepsilon_1$
for $n=1$. Here, $N$ denotes the $e$-folding number counted from the
initial time $N_i=0$. For $B=0.5964\,(< B_c)$, after the oscillation (in the
direction of $T_R$), it approaches to the exact de Sitter
solution. Meanwhile, for $B=0.59645, 0.5965\, (> B_c)$, after the
oscillation, the modulus field $T$ starts to roll down the potential slowly
all along $|T| \simeq 1$. After a sufficiently long $e$-folding, $T$
starts to oscillate again, terminating the slow-roll phase. In
Fig.~\ref{Fg:epsilon}, we show the result in case we set the initial condition as  
\begin{align}
 & \left( T_R,\, \frac{d T_R}{d N},\, T_I,\, \frac{d T_I}{d N} \right) =
 (0.9,\, 0.1,\, 0.0001,\, 0.0001)
\end{align}
at $N_i=0$. As expected from the fact that the
potential around $T \simeq 1$ is approximated by the hilltop
potential, which is small field, we need to fine tune the initial values of $T_I$ and 
$d T_I/dN$. (To have an inflationary solution, $|T_I| \ll 1$ and
$|dT_I/dN| \ll 1$ are preferred). We can vary the initial values of
$T_R$ and $d T_R/dN$ in a broader region, since $V$ takes a minimum
value at $T^m_1$ under the variation of $T_R$ around $T_R \simeq 1$ with $T_I$ fixed at $T_I \simeq
0$.

\begin{figure}[t]
\begin{center}
\begin{tabular}{cc}
\includegraphics[width=10cm]{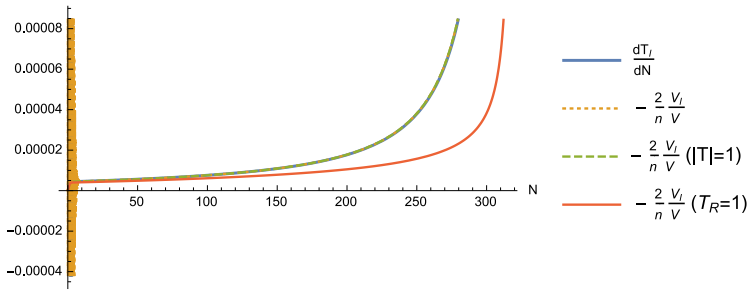}
\end{tabular}
\caption{We examine the validity of the slow-roll approximation and
 the assumptions $T_R \simeq 1$ and $|T|=1$.}
\label{Fg:trajectory}
\end{center}
\end{figure}

During the inflationary stage, the trajectory almost traces
the edge of the fundamental domain $|T|=1$, changing the angle
$\theta$ from $0$ towards $\pm \pi/6$. In this stage, the trajectory is
predominantly determined by the axion $T_I$ and $T_R$ almost stays $T_R \simeq 1$,
i.e., $|\dot{T}_R/\dot{T}_I| \ll 1$. Then, the non-canonical kinetic term does not affect the evolution much,
and the equation of motion is approximated by 
\begin{align}
 & \frac{d T_I}{dN} \simeq - \frac{2 T_R^2}{n} \frac{V_I}{V}  \simeq
 - \frac{2}{n} \frac{V_I}{V}\,. 
\end{align}
Indeed, as shown in Fig.~\ref{Fg:trajectory}, $d T_I/ d N$ (the blue solid line)
agrees with $- 2 V_I/(n V) $ (the orange dashed line) in a good precision. 

While the change of $T_R$ is tiny, this does not mean that we can
neglect the curvature of the trajectory on $|T| \simeq 1$ by setting $T_R=1$, because the
potential gradient in the direction of $T_R$ is large. Therefore, in
order to compute the accurate value of, say, $\frac{d T_I}{dN}$, we need
to take into account the variation of $V_I/V$ due to the tiny change of
$T_R$. As shown in Fig.~\ref{Fg:trajectory}, the value of $- (2 V_I/n V)$
evaluated for $T(N)= 1+ i T_I(N)$ (the red solid line) sizably deviates from the one
evaluated for the actual value of $T(N)$  (the orange dashed line). By contrast,
the value of $- (2 V_I/n V)$ evaluated for $T(N)=\sqrt{1- T_I^2(N)}+ i
T_I(N)$  (the green dashed line) almost agrees with the actual value, which indicates the importance to
take into account the curvature of the trajectory on $|T| \simeq 1$.

When $\theta$ approaches to $\theta = \pi/6$, the slow-roll evolution
abruptly ends. In this stage, the deviation from the canonical kinetic
term is onset and the trajectory depends on the two fields $T_R$ and
$T_I$. Therefore, in order to determine the total $e$-folding number, we
need to solve the two-field system with the non-canonical kinetic terms. Because of the non-linear velocity terms in
the equations of motion for $T_R$ and $T_I$, once $T_R$ and $T_I$
acquire non-negligible velocities, the slow-roll phase finishes much
faster than the case with the same scalar potential $V$, but with the
canonical kinetic term.

\begin{figure}[t]
\begin{center}
\begin{tabular}{cc}
\includegraphics[width=7.5cm]{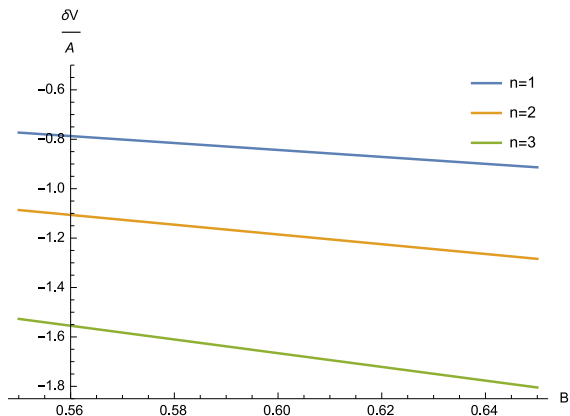}
\end{tabular}
\caption{This plot shows the value of $\delta V/A$ for each value of $B$
 and $n$.}
\label{Fg:dV}
\end{center}
\end{figure}
In Fig.~\ref{Fg:dV}, we plot the value of $\delta V/A = {\cal O}(\delta V/V)$ which was
introduced to set $V=0$ at the global minimum $T^m_2$ for 
$0.55 \leq B \leq 0.65$, which includes the range of $B$ where we found
the inflationary solution. Since $\delta V$ and $B$ are related as in
Eq.~(\ref{Exp:delta V}), $\delta V$ takes a negative value for $B> 0$.

\subsubsection{Case2: $B \leq  0$} \label{SSSec:Bnp}
For $B< 0$, $\delta V$ becomes positive and for $B=0$, $\delta V$
becomes 0. For $B< 0$, one may simply identify $\delta V$ with an explicit SUSY breaking uplift term. 
For $B<0$, $\theta=\pi/6$ becomes the local minimum or the
inflection point. We especially searched an inflationary solution which starts
around $T= e^{i \pi/6}$ and terminates around $T\simeq 1$. However, all
the solutions we found are either the one which does not embark an
inflationary trajectory ($- 3 \leq B \alt -0.8$) or the one which approaches to the
eternal de Sitter phase ($-0.8  \alt B < 0$). We leave an extensive
study of this case for a future study. For $B=0$, we did not find an
inflationary solution with a sufficiently large $e$-folding either.

As we have seen so far, requesting the modular invariance restricts a
possible form of the scalar potential. As $T_R$ becomes larger, the
scalar potential $V$ increases and for $T_R > 1$, the gradient of $V$
appears to be too steep to have an inflationary solution. The region where $V$ becomes nearly flat can be found only around $|T| \simeq 1$. 
An inflationary solution with a successful exit was found only in the
small parameter space of $B$, i.e, $B_c  \leq B \alt 0.6$ and it is
a small field inflation. We found that a large field
inflation is hardly realized in moduli inflation with the modular
invariance, which is crucially different from the axion
inflation models where the modular invariance is broken~\footnote{
Small-field axion inflation also can be realized.
See, e.g., Refs.~\cite{Peloso:2015dsa,Kobayashi:2015aaa}.}.

\section{Primordial perturbations}  \label{Sec:Primordial}
In the previous section, for $B_c  \leq B \alt 0.6$, we found a new
inflationary solution, whose gross potential during the slow-roll phase
is approximately given by the hilltop type potential. In this section, we compute the
primordial perturbations generated in this model and discuss the consistency
with the CMB measurements.

\subsection{Formulae of the primordial perturbations}
As discussed in the previous section, the trajectory during inflation is
mostly determined by the single degree of freedom $\theta$, the angular
direction. Following Gordon et al.~\cite{Gordon:2000hv}, we define the adiabatic
perturbation as the fluctuation in the direction of the background
trajectory:
\begin{align}
 & \delta \sigma = \frac{\dot{T}_R}{\dot{\sigma}}\, \delta
 T_R + \frac{\dot{T}_I}{\dot{\sigma}}\, \delta
 T_I \,
\end{align}
with
\begin{align}
 & \dot{\sigma} \equiv \sqrt{\dot{T}_R^2 + \dot{T}^2_I}\,.
\end{align}
Performing the time coordinate transformation, we obtain the relation
between the curvature perturbation on the slicing $\delta \sigma = 0$,
$\zeta$, and $\delta \sigma$ on the flat slicing, $\delta \sigma_f$,
as 
\begin{align}
    & \zeta  = - \frac{H}{\dot{\sigma}} \delta \sigma_f \,. 
\end{align}

During inflation, since $|\dot{T}_R/\dot{T}_I| \ll 1$, which implies
$\varepsilon_1 \simeq \varepsilon_{1, I}$, and $\delta T_R$ is
suppressed due to the large mass, the curvature perturbation $\zeta$ is
predominantly determined by the fluctuation of the axion field $T_I$ as
\begin{align}
 & \zeta \simeq - \frac{H}{\dot{T}_I} \delta T_{I,\, f}\,,
\end{align}
where $\delta T_{I,\, f}$ is the fluctuation of $T_I$ on the flat
slicing. Using this formula, we can compute the power spectrum of
$\zeta$ simply by quantizing the canonically normalized scalar field 
$$
 \sqrt{\frac{n}{2}} \frac{\delta T_{I,\,f}}{T_R}\,.
$$
The primordial spectrum of $\zeta$, defined by 
\begin{align}
 & {\cal P}_\zeta(k) \equiv \frac{k^3}{2 \pi^2} |\zeta_k|^2 \,,
\end{align}
is given by the standard formula 
\begin{align}
 & {\cal P}_\zeta(k) \simeq \frac{1}{8 \pi^2}
 \frac{H_k^2}{\varepsilon_{1,I,k}}  \simeq \frac{1}{8 \pi^2}  \frac{H_k^2}{\varepsilon_{1,k}}\,.
\end{align}
Here and hereafter, we put the index $k$ on background quantities which
are evaluated at the Hubble crossing time of the mode $k$. 
Then, the spectral index $n_s$ and the tensor to scalar ratio $r$ are given by the standard formulae as
\begin{align}
 & n_s- 1 \simeq - 2 \varepsilon_{1,k} - \varepsilon_{2,k}\,, \\
 & r \simeq 16 \varepsilon_{1, k}\,. 
\end{align}
During inflation, since $|\varepsilon_{1,\, k}| \ll |\varepsilon_{2,\, k}|$, the
spectral index is almost determined only by $\varepsilon_2$ at the Hubble crossing as 
\begin{align}
 & n_s- 1 \simeq - \varepsilon_{2,k} \,. 
\end{align}
As seen in this subsection, the primordial spectrum in our model is
superficially given by the same expression as the one for the canonical scalar field,
because they are determined during the slow-roll phase, when the
deviation from the canonical kinetic term is not effectively important.

\subsection{Results}
\begin{figure}[t]
\begin{center}
\begin{tabular}{cc}
\includegraphics[width=7.5cm]{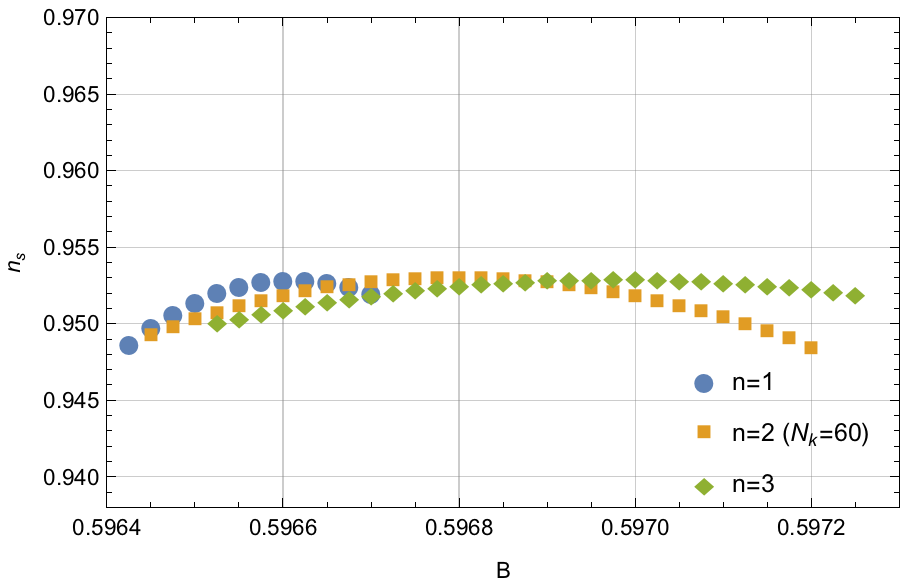}
\includegraphics[width=7.5cm]{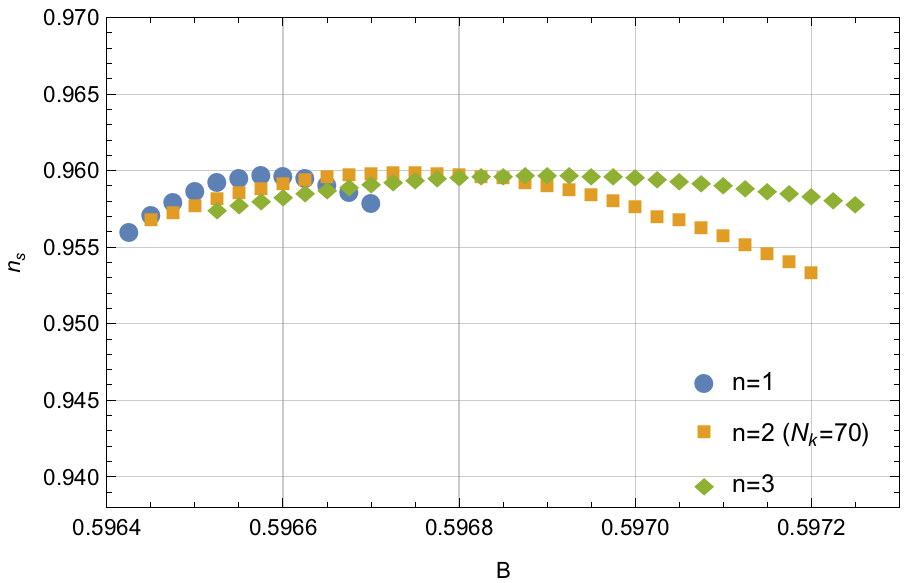}
\end{tabular}
\caption{These figures show the values of the scalar spectral index for
 different values of $B$ and $n=1,\, 2,\, 3$. The left panel is for
 $N_k=60$ and the right panel is for $N_k=70$.}
\label{Fg:ns}
\end{center}
\end{figure}

We compute the primordial spectrum for the values of $B$ which can yield
the inflationary solution ($B_c \alt B \alt 0.6 $). The result is presented
in Fig.~\ref{Fg:ns}. We determined the end of inflation as the time when
$\varepsilon_1$ becomes ${\cal O}(1)$ and $N_k$ denotes the $e$-folding
number from the Hubble crossing time of the mode $k$ to the end of inflation. 
As we increase $B$, at a certain value of $B$,
some of the slow-roll parameters $\varepsilon_m$ with $m \geq 1$ become larger than 1. In this paper, we
only analyze the parameter space of $B$ where all of $\varepsilon_m$
stay smaller than 1 during inflation, deferring the study of the case with the violation of the
slow-roll condition for our future publication~\cite{next}.  

As we decrease the value of $B$, the spectral index $n_s$ increases,
approaching to the scale invariant spectrum. However, at a certain
value of $B$, the spectral index starts to decrease. The value of $B$ at
which $n_s$ reaches the maximum, depends on $n$. For a larger value of $B$
(in the right sides of two panels of Fig.~\ref{Fg:ns}), the spectral
index $n_s$ is closer to 1 for $n=3$ than the one for $n=1$ and for smaller values (in the left sides), this becomes
opposite.

This behaviour can be better understood by
using the approximate expression of $V$ in the limit $|\theta| \ll 1$. Using
Eqs.~(\ref{Vht}) and (\ref{tempmu}), we find that the curvature of the potential for
the canonically normalized field is given by
\begin{align}
 &(-\varepsilon_2 \simeq)  \frac{4}{n}\, \frac{1}{V} \frac{d^2 V}{d T_I^2} \simeq - \frac{8}{n \mu^2(B)} \propto
 - \frac{B- B_c}{n^2}\,. \label{curvautreHT}
\end{align} 
As is mentioned previously, since $\varepsilon_1$ is much smaller than $\varepsilon_2$, $n_s-1$ is
determined by $- \varepsilon_2 \simeq (4/n) (d^2 V/d T_I^2)/V$. Therefore, if
the potential during inflation is all along given by the hilltop
potential (\ref{Vht}),  the deviation from the scale invariant spectrum
becomes smaller for a larger value of $n$ and for a smaller value of $B-B_c$.

Obviously, this does not fully explain our result, which has the maximum
value of $n_s$ and shows the more complicated dependence on $n$. In
addition, if the potential during inflation is accurately given by  
Eq.~(\ref{Vht}), the spectral index does not depend on $N_k$, i.e., the
spectral index can be determined without knowing when inflation
ends. By contrast, our result significantly depends on $N_k$. Because of
that, there should be a non-negligible deviation from the hilltop potential
(\ref{Vht}), while it is useful simply to understand the gross feature
of the scalar potential. In fact, while $\varepsilon_m$ with $m \geq 3$
become negligibly small in case the potential is precisely given by
Eq.~(\ref{Vht}), in this model, $\varepsilon_m$ with $m \geq 3$ take
non-vanishing values, letting $\varepsilon_2$ vary in time.

For a larger value of $B- B_c$ and a smaller value of $n$, the total
$e$-folding number becomes smaller (see Fig.~\ref{Fg:epsilon}) and the
value of $T_I(N_k)$ for a particular value of $N_k$ becomes smaller (as there was not enough time until
$N_k$ to proceed towards the global minimum).  Then, for a large value
of $B- B_c$ and a small value of $n$, $\theta \simeq T_I$ tends 
to explore a region with a more gentle potential gradient at each $N_k$, which
makes the spectrum closer to the scale-invariant spectrum.  The spectral
index $n_s$ is determined as a result of the competition of these two
effects. When the absolute value of the curvature of the gross potential given by
Eq.~(\ref{curvautreHT})  is small, the second effect is more important
and the deviation from the scale invariance becomes larger for a smaller
value of $B- B_c (>0)$ and a larger value of $n$. Thus, as presented in
Fig.~\ref{Fg:ns}, the deviation of the scale invariant spectrum $1-n_s$
has the minimum value in $B_c \alt B \alt 0.6$.

\begin{figure}[t]
\begin{center}
\begin{tabular}{cc}
\includegraphics[width=7.5cm]{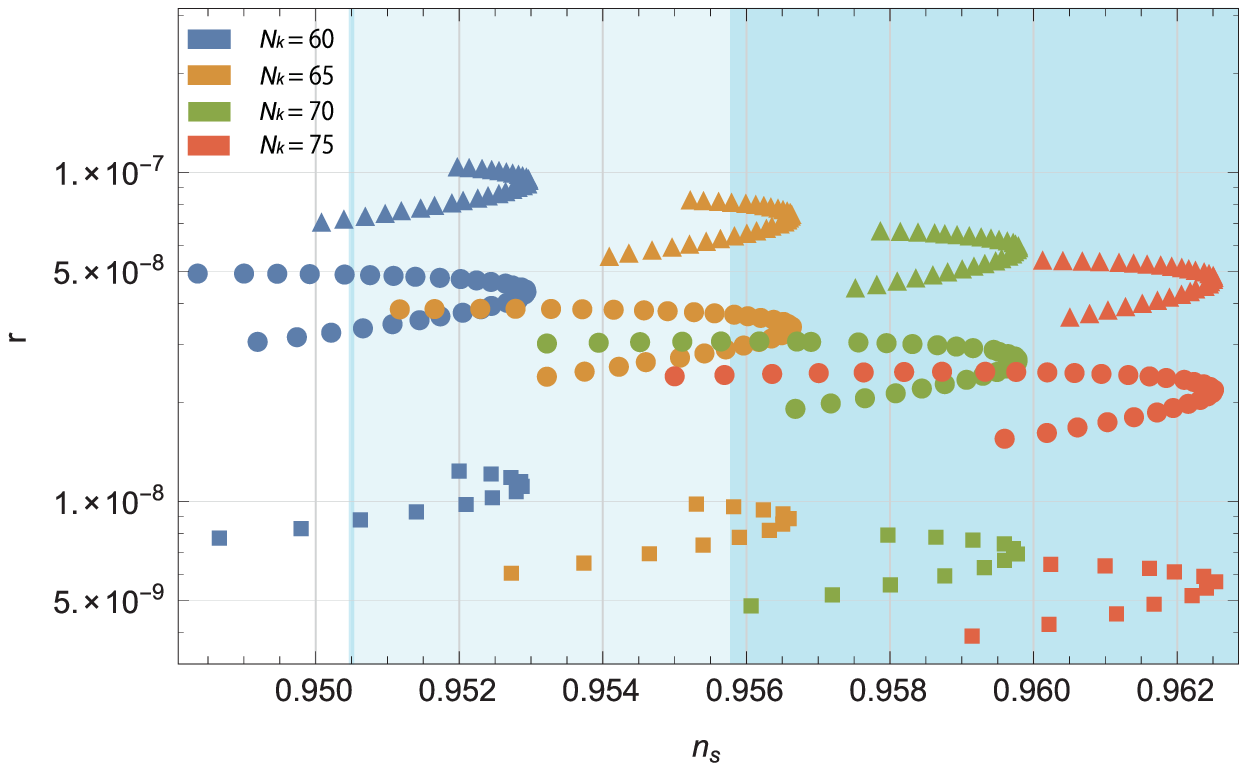}
\includegraphics[width=7.5cm]{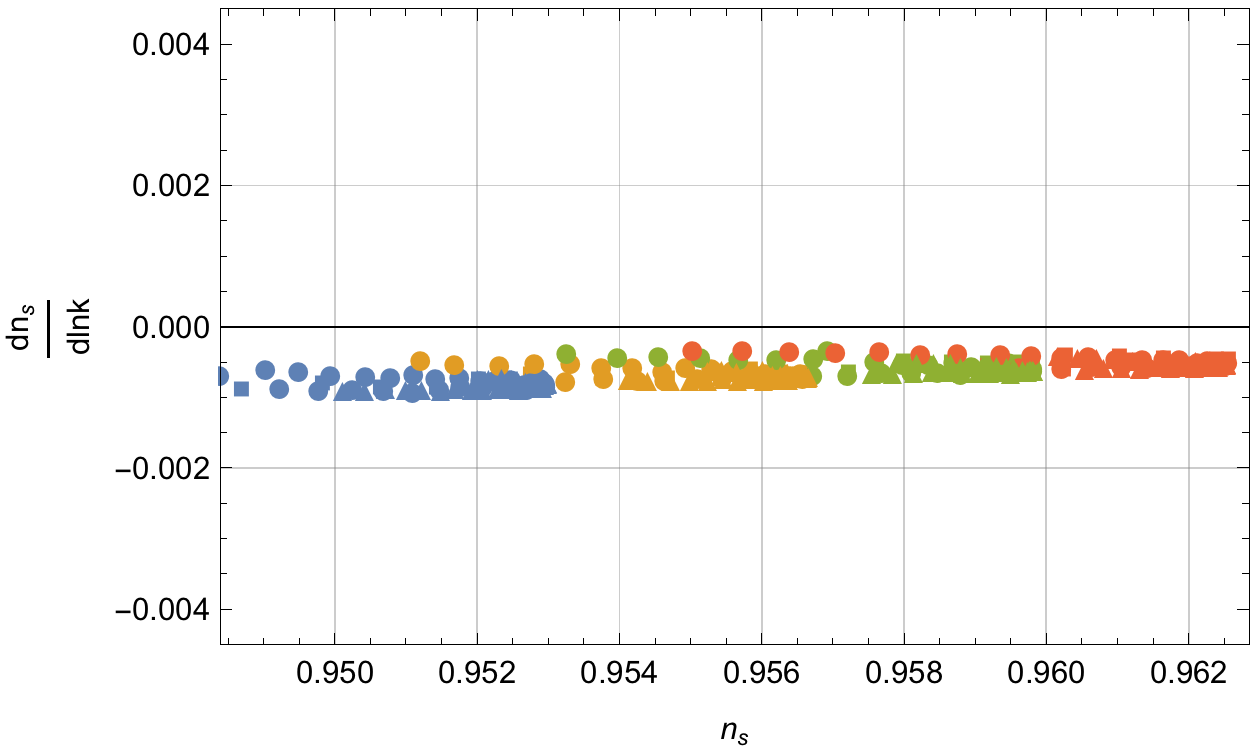}
\end{tabular}
\caption{The left panel shows the plot of the tilt and the tensor to
 scalar ratio and the right panel shows the plot of the tilt and the
 running. The dark blue region is compatible with the Planck
 measurement in 68\% CL and the light blue region is in 95\% CL~\cite{Planck15}. The blue markers are for $N_k=60$, the orange
 ones are for $N_k =65$, the green ones are for $N_k=70$, and the red
 ones are for $N_k=75$. The square markers are for $n=1$, the circle
 ones are for $n=2$, and the triangle ones are for $n=3$.  
As we increase $N_k$, more values of $B$ come to be observationally allowed.}
\label{Fg:nsr}
\end{center}
\end{figure}

The left panel of Fig.~\ref{Fg:nsr} shows the $(n_s, r)$-plot. The dark
 blue region is compatible with the Planck measurement in 68\% CL and
 the light blue region is in 95\% CL~\cite{Planck15}. Here, we 
marginalized the running $d n_s/d \ln k$. Since 
$\varepsilon_1 \simeq \varepsilon_{1,\, I}$ is ${\cal O}(10^{-8})$, the
upper bound on the tensor to scalar ratio $r$ does not place any meaningful constraint on the parameter space of $B$ and
$n$. The amplitude of the primordial gravitational waves is rather
small even among the small field models of inflation. This is because
$\varepsilon_1$ rapidly grows at the end of inflation due to the
non-canonical kinetic term. Therefore, the value of $\varepsilon_1$ for the CMB scales becomes much
smaller than the one for the model with the same scalar potential and
canonical kinetic term. In order to be compatible with the observations, the parameter $B$
should be fine tuned with 0.1 \% level accuracy of its value. This
may require a contrived setup in string theory, in particular in the
presence of non-negligible radiative corrections, while the fine
tuning problem is somehow common in small field models of inflation. 
In the right panel of Fig.~\ref{Fg:nsr}, we show the running of the
scalar spectral index. Since the slow-roll
parameters at the Hubble crossing are sufficiently small, there is no
enhancement of the running and the predicted value is obviously consistent with  
\begin{align}
 & \frac{d n_s}{d \ln k} = - 0.0057 \pm 0.0071 \qquad (68 \% CL,\, Planck {\rm TT},\, {\rm
 TE},\, {\rm EE},\,+ {\rm lowP})\,.
\end{align}

\begin{figure}[t]
\begin{center}
\begin{tabular}{cc}
\includegraphics[width=7.5cm]{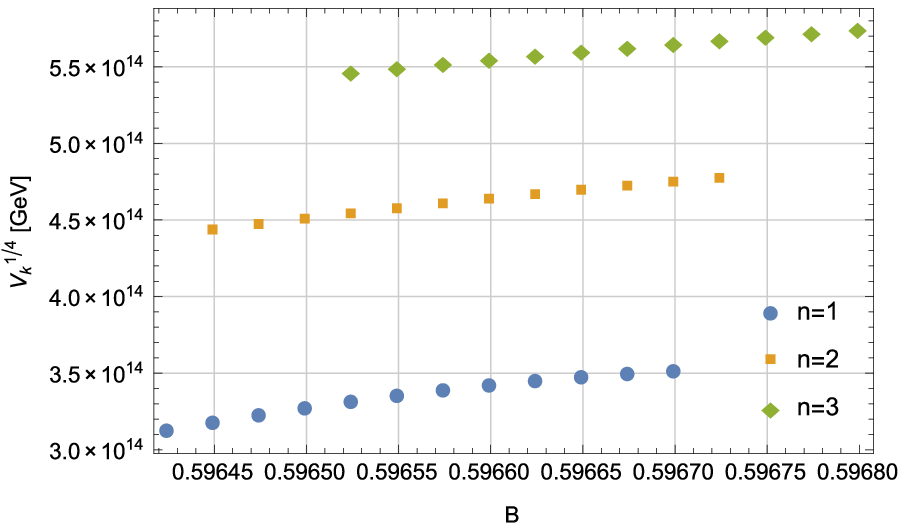}
\includegraphics[width=7.5cm]{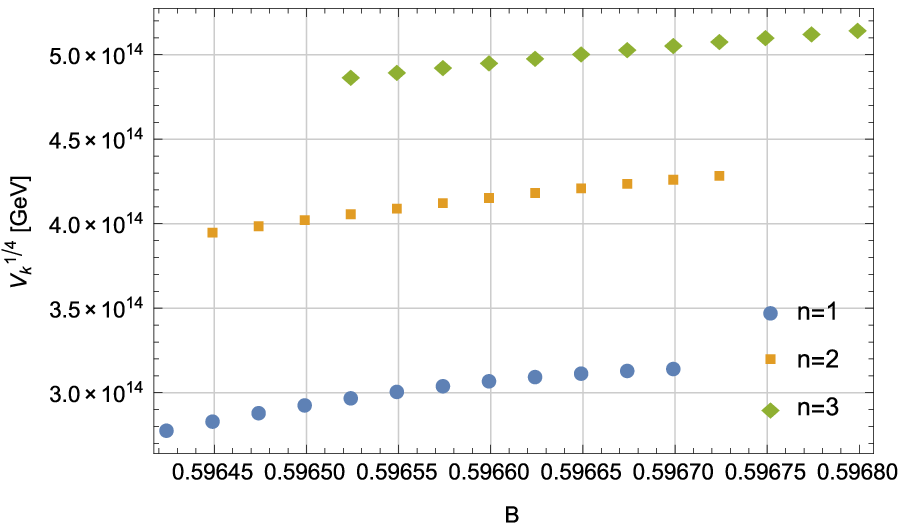}
\end{tabular}
\caption{The left panel shows the value of the scalar potential $V^{\frac{1}{4}}$[GeV] at the
 Hubble crossing for $N_k=60$ and the right panel shows the one for
 $N_k=70$. }
\label{Fg:VF}
\end{center}
\end{figure}
Inserting the value of the tensor to scalar ratio $r$ into
\begin{align}
 & V_k \simeq  \frac{3 \pi^2 A_s}{2}\, r\, M_{pl}^4 \simeq (1.88 \times 10^{16} {\rm
 GeV})^4\, \frac{r}{0.10}
\end{align}
where $A_s$ is the amplitude of the scalar power
spectrum~\cite{Planck15} , we obtain the value of the scalar potential
at the Hubble crossing, $V_k$. (See Fig.~\ref{Fg:VF}.) 
Since the tensor to scalar ratio $r$  is extremely small as 
$r={\cal O}(10^{-8})$, the scale of the scalar potential is of  
${\cal O}(10^{14} [{\rm GeV}])$, which is lower than the GUT scale. 
The Hubble scale during inflation is also low as ${\cal O}(10^{10})$[GeV].
Thus, the cosmological history is different from the one in large-field axion 
inflation with higher $V_k$ and $H$.

\begin{figure}[t]
\begin{center}
\begin{tabular}{cc}
\includegraphics[width=7.5cm]{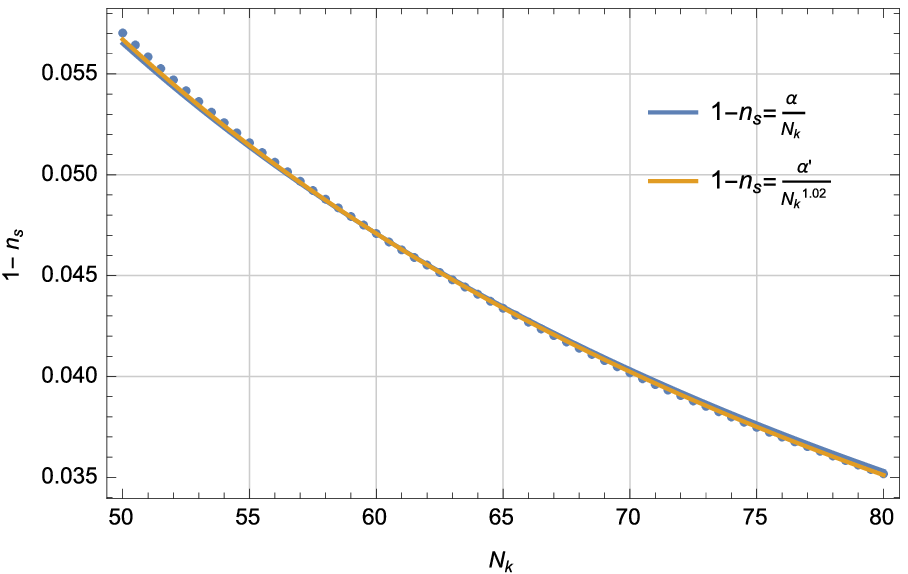}
\includegraphics[width=7.5cm]{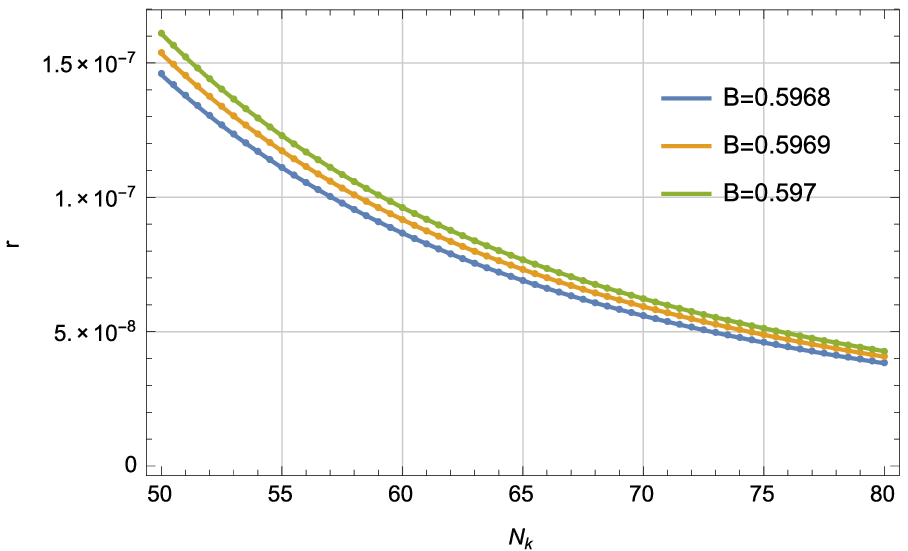}
\end{tabular}
\caption{The left panel shows the $N_k-(n_s-1)$ plot for $n=3$ and
 $B=0.5968$ and it can be well fitted by $n_s - 1 \simeq - \alpha/N_k$
 with $\alpha= 2.83$. (Slightly better for $n_s - 1 \propto
 1/N_k^{1.02}$.) The right panel shows that the $N_k-r$ plot is almost
 on the line of $r \simeq \frac{16}{\beta} N_k^{-\alpha}$.}
\label{Fg:nsNk}
\end{center}
\end{figure}

As was discussed in Refs.~\cite{Roest:2013fha, Creminelli:2014nqa}, the CMB observations suggest that
the spectral index $n_s$ is related to $N_k$ as 
\begin{align}
 & n_s - 1 = - \frac{\alpha}{N_k}\,. \label{nsNk}
\end{align}
In such a case, solving
\begin{align}
 & n_s-1 \simeq - \varepsilon_1(N_k) + \frac{d \ln \varepsilon_1(N_k)}{d
 N_k} \simeq \frac{d \ln \varepsilon_1(N_k)}{d
 N_k}  \,, \label{nsNk2}
\end{align}
we can compute $\varepsilon_1(N_k)$ and the tensor to scalar
ratio. Figure \ref{Fg:nsNk} shows that when the parameter $B$ is set to the
ones which are compatible with the Planck measurements, the spectral
index in this model is given by Eq.~(\ref{nsNk}) in a good
accuracy. For instance, for $n=3$ and $B=0.5968,\,0.5969, \, 0.597$,
$\alpha$ is given by $\alpha=2.85,\, 2.83,\, 2.82$, respectively.   
Then, using Eqs.~(\ref{nsNk}) and (\ref{nsNk2}) and integrating about
$N_k$, we obtain $r$ as  
\begin{align}
 & r \simeq \frac{16}{\beta} N_k^{-\alpha}\,, 
\end{align}
where $\beta$ is an integration constant. With an appropriate choice of $\beta$,
this expression reproduces the numerical result of $r$ as it should be (see the
right panel of Fig.~\ref{Fg:nsNk}).

\subsection{Reheating process and estimation of $N_k$}
The presence of the minimum value of $1-n_s$ is very important, because it
gives a lower bound on $N_k$ to be compatible with the observations. 
The $e$-folding $N_k$ should depend on the various physics after inflation and is given 
by~\cite{Planck15, LL}
\begin{align}
 & N_k  \approx 66 - \ln \left( \frac{k}{a_0 H_0} \right) + \frac{1}{4}
 \ln \left( \frac{V_k^2}{M_{pl}^2\, \rho_{end}} \right) + \frac{1- 3
 w_r}{12(1+ w_r)} \ln \left( \frac{\rho_{th}}{\rho_{end}} \right)\,,
\end{align}
where $a_0$ and $H_0$ are the scale factor and the Hubble parameter
at present and $\rho_{end}$ and $\rho_{th}$ are the energy densities at
the end of inflation and at the time when the universe was thermalized after
inflation, respectively. The parameter $w_r \equiv p_r/\rho_r$
characterizes the effective equation of state during the oscillatory
phase after inflation. When we insert the typical value of $V_k$ in
this model and $V_k \simeq \rho_{end}$ into the above expression, we obtain
\begin{align}
 & N_k  \approx 58 - \ln \left( \frac{k}{a_0 H_0} \right) + \frac{1- 3
 w_r}{12(1+ w_r)} \ln \left( \frac{\rho_{th}}{\rho_{end}} \right)\,.
\end{align}
Therefore, in particular, when the thermalization process is completed
instantaneously, giving $\rho_{th} \simeq \rho_{end}$, this model is almost
marginal at 95\% CL.

Meanwhile, when $\rho_{th} < \rho_{end}$ and $w_r < 1/3$, a smaller
value of $N_k$ will be predicted, which makes more unlikely to be
compatible with the Planck data. In a single field model with the canonical kinetic term and a power-law scalar
potential, $w_r > 1/3$ requires the power of the scalar potential should
be larger than 4. In our setup, the computation of $w_r$ will be more
complicated because the evolution of the oscillatory phase is determined
by the coupled two fields $T_R$ and $T_I$ with the non-canonical kinetic
term. To calculate $\rho_{th}$, we need to know the coupling between the
inflaton and the standard model sector. In this model, since the
inflaton is the modulus field, whose coupling to the standard model
sector can be identified, $\rho_{th}$ will be predictable. Thus, by combining
the information about the reheating, it may be possible to testify this model. A further
study will be reported in our forthcoming publication~\cite{next}.

\section{Conclusion}  \label{Sec:Conclusion}

In this paper, we investigated a phenomenological consequence of the
modular invariance. In particular, we studied the moduli inflation with the Lagrangian
which preserves the modular invariance. In this model, the minimum of
the scalar potential, is located at the edge of the fundamental region for the explored values of
$B$. Setting the scalar potential at the global
minimum to 0, we find that the Lagrangian density is determined only by
the three parameters $A$, $B$, and $n=1,\, 2,\, 3$. The modular invariance
constrains the shape of the scalar potential (which is mainly determined
by $B$) and the region with the flat potential was found only around $|T| =1$.

A successful inflation model was found only in the restricted parameter
region of $B$, $B_c < B \alt 0.6$. 
The background trajectory during the slow-roll phase is almost
determined by the axion, the imaginary part of the modulus field. Unlike
in most of the successful axion inflation models, whose trajectory traces the large
field periodic scalar potential, the scalar potential in this model is
given by the hilltop potential $V_{ht}$ with the small modulation, i.e., small field inflation. Then, the slow-roll inflation can take place
without the super-Planckian excursion as in the natural inflation,
which may give rise to a conflict to the UV completion of quantum gravity.

The predicted primordial gravitational waves are too tiny to be able to
detect them in a near future. Then, it is unlikely to falsify this
model only by examining the primordial perturbations. However, since the
coupling of the modulus field to the standard model sector can be
determined, the reheating temperature and the equation of state during
inflation are potentially calculable in a self-complete way. Therefore,
by studying possible values of $N_k$, it may be possible to further
examine the observational consistency of this this model. This result
will be reported in our future study~\cite{next}.

In this paper, we considered the simplest form of
the superpotential which preserves the modular invariance. In
Ref.~\cite{Cvetic:1991qm}, a generalization of the modular invariant model was
explored. It may be interesting to see if this generalization allows us
to find a new type of inflation. In the generalized setup, one may want to look for an inflation model
where the global minimum can be set to $V=0$ by introducing a positive
$\delta V$, since in such a case, we can identify $\delta V$ with an explicit SUSY breaking uplift term. 
This investigation is left for a future
study.

\section*{Acknowledgements}
We would like to thank T.~Higaki, K.~Ichiki, and H.~Otsuka for useful
discussions. This project was initiated and completed under the support
of Building of Consortia for the Development of Human Resources in
Science and Technology. T.~K. was supported in part by the Grant-in-Aid
for Scientific Research No.~25400252 and No.~26247042 from the Ministry
of Education, Culture, Sports, Science and Technology (MEXT) in
Japan. D.~N. is supported in part by MEXT Grant-in-Aid for Scientific Research on Innovation Areas, Nos. 15H05890. Y.~U. is supported by JSPS Grant-in-Aid for Research Activity
Start-up under Contract No. 26887018.


\begin{thebibliography}{99}



\bibitem{Hamidi:1986vh} 
  S.~Hamidi and C.~Vafa,
  Nucl.\ Phys.\ B {\bf 279}, 465 (1987);
  L.~J.~Dixon, D.~Friedan, E.~J.~Martinec and S.~H.~Shenker,
  Nucl.\ Phys.\ B {\bf 282}, 13 (1987);
%
  T.~T.~Burwick, R.~K.~Kaiser and H.~F.~Muller,
  Nucl.\ Phys.\ B {\bf 355}, 689 (1991);
%
  J.~Erler, D.~Jungnickel, M.~Spalinski and S.~Stieberger,
  Nucl.\ Phys.\ B {\bf 397}, 379 (1993)
  [hep-th/9207049];
%
  K.~-S.~Choi and T.~Kobayashi,
  Nucl.\ Phys.\ B {\bf 797}, 295 (2008)
  [arXiv:0711.4894 [hep-th]].
%

\bibitem{Cvetic:2003ch} 
  M.~Cvetic and I.~Papadimitriou,
  Phys.\ Rev.\ D {\bf 68}, 046001 (2003)
  [Erratum-ibid.\ D {\bf 70}, 029903 (2004)]
  [hep-th/0303083];
  S.~A.~Abel and A.~W.~Owen,
  Nucl.\ Phys.\ B {\bf 663}, 197 (2003)
  [hep-th/0303124];
%
  D.~Cremades, L.~E.~Ibanez and F.~Marchesano,
  JHEP {\bf 0307}, 038 (2003)
  [hep-th/0302105];
  S.~A.~Abel and A.~W.~Owen,
  Nucl.\ Phys.\ B {\bf 682}, 183 (2004)
  [hep-th/0310257].

\bibitem{Cremades:2004wa}
  D.~Cremades, L.~E.~Ibanez and F.~Marchesano,
  JHEP {\bf 0405}, 079 (2004)
  [hep-th/0404229];
%


\bibitem{Ibanez:2012zz}
  L.~E.~Ibanez and A.~M.~Uranga,
  ``String theory and particle physics: An introduction to string phenomenology,''
  Cambridge University Press (2012).


  H.~Abe, K.~S.~Choi, T.~Kobayashi and H.~Ohki,
  JHEP {\bf 0906}, 080 (2009)
  [arXiv:0903.3800 [hep-th]].


\bibitem{Kaplunovsky:1987rp} 
  V.~S.~Kaplunovsky,
  Nucl.\ Phys.\ B {\bf 307}, 145 (1988)
  [Erratum-ibid.\ B {\bf 382}, 436 (1992)]
  [hep-th/9205068].


\bibitem{Dixon:1990pc} 
  L.~J.~Dixon, V.~Kaplunovsky and J.~Louis,
  Nucl.\ Phys.\ B {\bf 355}, 649 (1991).



\bibitem{Derendinger:1991hq} 
  J.~P.~Derendinger, S.~Ferrara, C.~Kounnas and F.~Zwirner,
  Nucl.\ Phys.\ B {\bf 372}, 145 (1992).

\bibitem{Ferrara:1990ei}
  S.~Ferrara, N.~Magnoli, T.~R.~Taylor and G.~Veneziano,
  Phys.\ Lett.\ B {\bf 245}, 409 (1990).
  doi:10.1016/0370-2693(90)90666-T

\bibitem{Cvetic:1991qm}
  M.~Cvetic, A.~Font, L.~E.~Ibanez, D.~Lust and F.~Quevedo,
  Nucl.\ Phys.\ B {\bf 361}, 194 (1991).
  doi:10.1016/0550-3213(91)90622-5

\bibitem{Abe:2014xja}
  H.~Abe, T.~Kobayashi and H.~Otsuka,
  JHEP {\bf 1504} (2015) 160
  [arXiv:1411.4768 [hep-th]].

\bibitem{Higaki:2015kta}
  T.~Higaki and F.~Takahashi,
  JHEP {\bf 1503}, 129 (2015)
  doi:10.1007/JHEP03(2015)129
  [arXiv:1501.02354 [hep-ph]].

\bibitem{Kappl:2015esy}
  R.~Kappl, H.~P.~Nilles and M.~W.~Winkler,
  Phys.\ Lett.\ B {\bf 753}, 653 (2016)
  doi:10.1016/j.physletb.2015.12.073
  [arXiv:1511.05560 [hep-th]].

\bibitem{Planck15}
  P.~A.~R.~Ade {\it et al.} [Planck Collaboration],
  arXiv:1502.02114 [astro-ph.CO].

\bibitem{Freese:1990rb} 
  K.~Freese, J.~A.~Frieman and A.~V.~Olinto,
  Phys.\ Rev.\ Lett.\  {\bf 65}, 3233 (1990).

\bibitem{Pajer:2013fsa}
  E.~Pajer and M.~Peloso,
  Class.\ Quant.\ Grav.\  {\bf 30}, 214002 (2013)
  doi:10.1088/0264-9381/30/21/214002
  [arXiv:1305.3557 [hep-th]].

\bibitem{Svrcek:2006yi} 
  P.~Svrcek and E.~Witten,
  JHEP {\bf 0606}, 051 (2006)
  doi:10.1088/1126-6708/2006/06/051
  [hep-th/0605206].


\bibitem{Kim:2004rp}
  J.~E.~Kim, H.~P.~Nilles and M.~Peloso,
  JCAP {\bf 0501} (2005) 005
  [hep-ph/0409138].
  

\bibitem{Abe:2014pwa}
  H.~Abe, T.~Kobayashi and H.~Otsuka,
  PTEP {\bf 2015} 6,  063E02
  [arXiv:1409.8436 [hep-th]].
  

\bibitem{Copeland:1994vg}
  E.~J.~Copeland, A.~R.~Liddle, D.~H.~Lyth, E.~D.~Stewart and D.~Wands,
  Phys.\ Rev.\ D {\bf 49}, 6410 (1994)
  doi:10.1103/PhysRevD.49.6410
  [astro-ph/9401011].

\bibitem{Choi:1994xg} 
  K.~Choi, J.~E.~Kim and H.~P.~Nilles,
  Phys.\ Rev.\ Lett.\  {\bf 73}, 1758 (1994)
  doi:10.1103/PhysRevLett.73.1758
  [hep-ph/9404311].









\bibitem{Peloso:2015dsa}
  M.~Peloso and C.~Unal,
  JCAP {\bf 1506}, no. 06, 040 (2015)
  doi:10.1088/1475-7516/2015/06/040
  [arXiv:1504.02784 [astro-ph.CO]].


\bibitem{Kobayashi:2015aaa}
  T.~Kobayashi, A.~Oikawa and H.~Otsuka,
  arXiv:1510.08768 [hep-ph].


\bibitem{Gordon:2000hv}
  C.~Gordon, D.~Wands, B.~A.~Bassett and R.~Maartens,
  Phys.\ Rev.\ D {\bf 63}, 023506 (2001)
  doi:10.1103/PhysRevD.63.023506
  [astro-ph/0009131].




\bibitem{next}
  R.~Iida,~T.~Kobayashi,~D.~Nitta,~and Y.~Urakawa,~in preparation.
 

\bibitem{Roest:2013fha}
  D.~Roest,
  JCAP {\bf 1401}, 007 (2014)
  doi:10.1088/1475-7516/2014/01/007
  [arXiv:1309.1285 [hep-th]].


\bibitem{Creminelli:2014nqa}
  P.~Creminelli, S.~Dubovsky, D.~Lopez Nacir, M.~Simonovic, G.~Trevisan, G.~Villadoro and M.~Zaldarriaga,
  Phys.\ Rev.\ D {\bf 92}, no. 12, 123528 (2015)
  doi:10.1103/PhysRevD.92.123528
  [arXiv:1412.0678 [astro-ph.CO]].

\bibitem{LL} 
  A.~R.~Liddle and S.~M.~Leach,
  Phys.\ Rev.\ D {\bf 68}, 103503 (2003)
  doi:10.1103/PhysRevD.68.103503
  [astro-ph/0305263].



\end{thebibliography}
\end{document}